\documentstyle{l-aa}
\input{psfig}

\font\tenbi=cmmib10
\newfam\bifam  \textfont\bifam=\tenbi

\def\etal{\mbox{et al.~}}

\begin{document}
 
\thesaurus{06(03.13.1; 03.13.2; 03.13.6; 06.15.1)}

\title{The art of fitting p-mode spectra: Part I. Maximum Likelihood Estimation}
\author{Thierry~Appourchaux\inst{1}, Laurent~Gizon\inst{1,2}, 
Maria-Cristina Rabello-Soares\inst{1}}
\institute{Space Science Department of ESA, ESTEC, NL-2200 AG 
Noordwijk
\and
W.W.Hansen Experimental Physics Laboratory, Center for Space Science and 
Astrophysics, Stanford University, Stanford, CA 94305-4085, USA}
\offprints{thierrya\mbox{@}so.estec.esa.nl}

\date{Received / Accepted}

\maketitle
\markboth{Appourchaux \etal: The art of fitting, Part I}{}
\begin{abstract}
In this article we present our state of the art of fitting 
helioseismic p-mode spectra.
We give a step by step recipe for fitting the spectra: statistics of the 
spectra both for spatially unresolved and resolved data, the use of Maximum 
Likelihood estimates, the statistics of the p-mode parameters, the use
of Monte-Carlo simulation and the significance of fitted parameters.  
The recipe is applied to synthetic low-resolution data, similar to 
those of the LOI, using Monte-Carlo simulations.  For such spatially resolved data, 
the statistics of the Fourier spectrum is assumed to be a multi-normal 
distribution; the statistics of the power spectrum is \emph{not} a $\chi^{2}$ with 
2 degrees of freedom.  Results for $l=1$ shows 
that all parameters describing the p modes can be obtained without bias 
and with minimum variance provided that the leakage matrix is known.  
Systematic errors due to an imperfect knowledge of the leakage matrix 
are derived for all the p-mode parameters.
\end{abstract}

\section{Introduction}
In the past decade, helioseismology has been able to provide the internal structure 
of the Sun and its dynamics.  These inferences have been made possible 
by inverting the frequencies and rotational splitting of the pressure modes.  The 
most commonly used technique for obtaining the p-mode parameters is 
to fit the p-mode spectra using Maximum Likelihood Estimators (MLE) 
assuming that the statistical distribution of the p modes in the power spectra is a 
$\chi^{2}$ with 2 degrees of freedom (Woodard 1984).  
The MLE with this statistics were first applied on helioseismology data 
by Duvall and Harvey (1986) and Anderson \etal (1990).  
This technique is used for fitting spectra obtained with integrated 
sunlight instruments.  For low- or high-resolution instruments, the
 $(m,\nu)$ power spectra are commonly fitted assuming that each 
 $m$ spectrum has the same statistics as the for the integrated 
 sunlight instruments (LOI 
 instrument: (Appourchaux \etal 1995; Rabello-Soares \etal 1997); GONG 
 instrument (Hill \etal 1996)).  
Unfortunately, none 
of these implementations are correct since the assumed statistics is wrong.  
Only Schou (1992) described a more correct way of fitting $(m,\nu)$ 
diagrammes using \emph{not} the power spectra 
but the complex Fourier spectra.

The pioneering work of Schou (1992) has inspired this series of 3 
articles for addressing our state of the art of fitting $(m,\nu)$ 
diagrammes.  In this paper (Part I), we describe the statistics of the p modes, and 
how the MLE can be used in helioseismology.  In Appourchaux \etal 
(1997)
(hereafter Part II), we 
show how one can measure the mode leakage matrix and the noise 
correlation from the data which knowledge is required for using the Part 
I.  In Appourchaux and Gizon (1998) (hereafter Part III), we applied these techniques 
to the LOI instrument of VIRGO on board SOHO (For a 
description of the performance of the instrument see Appourchaux \etal 
1997).
 
In this paper, we explain how the MLE can be used in helioseismology.  In the first section, we recall the properties of MLE.  
In the second 
section, we describe the statistics of the p-mode Fourier spectra.  In 
this section, we have generalized the approach of Schou (1992), to any 
complex leakage matrices.  We have also used complex matrices to 
generate the covariance matrices of the p modes and of the noise.  In the 
third section we show how to use Monte-Carlo simulations for testing 
both the use of MLE and the model of the p-mode spectra, and then 
conclude.

\section{Maximum Likelihood Estimators}
Some of the properties of MLE were given by Toutain and Appourchaux 
(1994).  We repeat them here for completeness.  We also address 2 issues 
that were not covered in their
paper: are MLE biased?, and how significant are the estimated 
parameters.

\subsection{Fundamental properties}
The aim of this section is to introduce some definitions and properties of  
MLE.  A comprehensive study of this area of statistics can be found, e.g. 
in Kendall and Stuart (1967). 
Given a random variable $x$ with a probability distribution $f(x,\vec{\lambda})$, 
where $\vec{\lambda}$ is a vector of $p$ parameters.
We define the logarithmic likelihood function $\ell$ of $N$ independent 
measurements $x_{k}$ of $x$ 
as
\begin{eqnarray}
\ln{L}=\ell = -\sum_{k=1}^{N} \ln f(x_k,\vec{\lambda}).
\end{eqnarray}
where $L$ is the likelihood.  The main property of $\ell$ is that the position of its minimum in 
the $\vec{\lambda}$-space gives an estimate of the most likely value of $\vec{\lambda}$,
 denoted hereafter as $\tilde{\vec{\lambda}}$. Hence $\tilde{\vec{\lambda}}$ is the solution of
the set of $p$ simultaneous equations :
\begin{eqnarray}
{{\partial \ell} \over{\partial {\lambda_i}}}=0 &with&i = 1, 2, ..., p.
\end{eqnarray}
Moreover, in the limit of very large sample ($N\rightarrow\infty$) this 
estimator 
$\tilde{\vec{\lambda}}$ 
tends to have a multi-normal probability distribution.  In this case, this 
estimator is asymptotically unbiased with minimum variance; which implies its 
expectation and variance are respectively:
\begin{eqnarray}
\lim_{N\rightarrow\infty} E(\tilde{\vec{\lambda}}) &=& \vec{\lambda}.\\
\lim_{N\rightarrow\infty} \sigma^{2}(\tilde{\vec{\lambda}}) &=& c_{ii}
\label{large}
\end{eqnarray}
where $c_{ii}$ are the diagonal elements of the inverse of the Hessian 
matrix $h$, with elements:
\begin{eqnarray}
h_{ij}=E({{\partial \ell}^{2} \over{\partial {\lambda_i}\partial {\lambda_j}}}).
\label{hij}
\end{eqnarray}
The covariances between any 2 components of $\tilde{\vec{\lambda}}$ are given by 
the corresponding off-diagonal elements of the inverse matrix.  Equation 
(\ref {hij}) is used when computing the so-called formal error bars on 
$\tilde{\vec{\lambda}}$; as a matter of fact according to the Cramer-Rao theorem, 
Eq. (\ref {hij}) gives only a lower bound to the 
error bars (Kendall and Stuart, 1967, reference therein).  Toutain and Appourchaux (1994) 
showed that Eq. (\ref {hij}) is valid for most purpose in 
helioseismology.

\subsection{Biased or unbiased?}
The fact that MLE are asymptotically unbiased does not necessarily mean that 
this property is kept for a finite amount of data.  As an 
example, it is well known that an estimator of the standard deviation ($\sigma$) 
of $N$ measurement of a normally distributed random variable $x$ is 
given by:
 \begin{equation}
 	\sigma^{2}=\frac{1}{N-1} \sum_{i=1}^{N} (x_{i}-\tilde{m})^{2}
 	\label{bias}
 \end{equation} 
where $x_{i}$ is the $i$-th measurement of the random variable $x$ and $\tilde{m}$ is an estimate 
of the mean.  It is well known that the $\sigma$ of Eq. 
(\ref {bias}) is unbiased.  In this case, MLE would give the following 
estimator:
 \begin{equation}
 	\sigma_{MLE}^{2}=\frac{N-1}{N} \sigma^{2}
 	\label{biasMLE}
 \end{equation}
 
 Clearly the MLE expression give a bias that vanish asymptotically for 
 an infinite number of points.  It is often difficult to derive 
 explicit relation, similar to Eq. (\ref {biasMLE}) between the estimator 
 and  the finite number of data points.  When analytical expression 
 can not be found, we advice to use Monte-Carlo simulations to verify 
 the unbiasness; an example for $l$=1 splittings is given in Chang 
 (1996) and Appourchaux \etal (1997). 
  
 In any case MLE are intrinsically biased estimators because they are 
 also minimum variance estimators (Kendall and Stuart, 1967).  It may be useful to find other 
 estimators that do not bias the estimates (Quenouille, 1956); they 
 might not necessarily have minimum variance.  These estimators are yet to be found.

\subsection{Significance of fitted parameters}
When one uses Least Square for fitting data, one can test the 
significance of its fitted parameters using the so-called $R$ test 
(Frieden, 1983).
For MLE, a useful test can be used: the likelihood ratio test.
It was first used by Appourchaux \etal (1994).  
This method requires to maximize the likelihood $e^{-\ell(\omega_{p})}$ of a given event 
where $p$ parameters are used to described the line profile.  If one wants to 
describe the same event with $n$ additional parameters, the likelihood 
$e^{-\ell(\Omega_{p+n})}$ 
will have to be maximized.  The likelihood ratio test consists in making the ratio 
of the two likelihood (Brownlee, 1965).  Using the logarithmic 
likelihood, we can define the ratio $\Lambda$ as:
\begin{equation}
\ln(\Lambda)=\ell(\Omega_{p+n})-\ell(\omega_{p})
\label{ll}
\end{equation}
If $\Lambda$ is close to 1, it means that there is no improvement in the maximized 
likelihood and that the additional parameters are not significant.  On the other 
hand, if $\Lambda \ll 1$, it means that $\ell(\Omega_{p+n}) \ll \ell(\omega_{p})$ and that 
the additional parameters are very significant.  Wilks (1938) 
 showed that for large sample size the distribution of $-2$ln$\Lambda$ tends to the
  $\chi^{2}(n)$ distribution.

\section{The statistics of p-mode spectra}
\subsection{Single mode}
It is well known that p modes are stochastically excited oscillators 
(Kumar \etal, 1988).  The source of excitation lies in the many 
granules covering the Sun.  The modes are assumed to be independently excited 
provided that their spatial scale is larger than the granule size 
(Chang, 1996).  From the equation of an oscillator, the 
statistics of the p-mode profile can be derived as:

 \begin{equation}
 	\frac{d^{2}x}{dt^{2}}+2\pi\gamma\frac{dx}{dt}+(2\pi)^{2}\nu_{0}^{2}x=F(t)
 	\label{a}
 \end{equation} 
where $t$ is the time, $x$ is the displacement, $\gamma$ is the damping 
term or the linewidth, 
$\nu_{0}$ is the frequency of the mode and $F(t)$ is the forcing 
function.  From this equation the Fourier transform of $x$ can be 
written as:

 \begin{equation}
 	\tilde{x}(\nu)=\frac{\tilde{F}(\nu)}{(2\pi)^{2}(\nu_{0}^{2}-\nu^{2}
 	+i\gamma\nu)}
 	\label{b}
 \end{equation} 
where $\tilde{x}(\nu)$ and $\tilde{F}(\nu)$ are the Fourier transform 
of $x(t)$ and $F(t)$.  From the large number of granules, it can be derived that the forcing function 
is normally distributed.  Therefore the 2 components (the real and imaginary 
parts) of the Fourier transform of the forcing function are also normally distributed.  For the p 
modes, each component of the Fourier transform is normally distributed 
with a mean of zero, and a variance given by:
 \begin{equation}	
 	\sigma^{2}(\nu)=\frac{1}{2}\frac{\sigma^{2}_{\tilde{F}}(\nu)}
 	{(2\pi)^{4}[(\nu_{0}^{2}-\nu^{2})^{2}+\nu^{2}\gamma^{2}]}
 	\label{c}
 \end{equation} 
The square of the modulus of $\tilde{x}(\nu)$, or 
power spectrum, has a $\chi^{2}$ with 2 degree of freedom statistics 
and its mean is given by Eq. (\ref{c}).
This is the p-mode profile that is usually approximated by a 
Lorentzian profile.  Similarly other effects such as asymmetry can be 
introduced in the profile of Eq. (\ref {c}).

\subsection{Unresolved observations}
Instruments integrating over the solar surface the velocity or the 
intensity signal observe a superposition of various modes of 
different degrees.  They are mainly sensitive to the low-degree 
modes ($l \le 4$).  For a 
given $l$, they detect a mixing of azimuthal order $m$ for which a 
visibility is prescribed (Toutain and Gouttebroze, 1994; 
Christensen-Dalsgaard and Gough, 1982).  Most often they can only detect modes for 
which $l+m$ is even.  Since the Fourier components of the observed 
time series have a normal 
distribution, and since the different $m$ are uncorrelated, the statistics of the power 
spectra of unresolved observation is a $\chi^{2}$ with 2 degrees of 
freedom.  Toutain and Appourchaux (1994) gave an analysis of 
the problem associated with these observations; we will not repeat it 
here.

\subsection{Resolved observations}
When the solar image is resolved many more degrees can be detected 
making the data analysis somewhat more complicated.  In order to 
extract a single $l,m$ mode from resolved observations, one has to apply a 
specific spatial filter or weight to the velocity or intensity 
images.  Most often these weights are such that imperfect 
isolation of the $l,m$ mode is achieved; especially because the most 
commonly used filters (spherical harmonics) are \emph{not} orthogonal over 
half a sphere.  This leads to the existence of 
other modes in the Fourier spectrum generated for a given $l,m$ filter.  
Therefore, the observed 
Fourier spectrum is a linear combination of the modes 
to be detected.  This linear combination of the modes can be understood 
as modes leaking into each other spectrum: this is represented by the so-called 
\emph{leakage} matrix.  These leakages will produce correlations 
between the different Fourier spectra.  These correlations will modify 
the statistics of the Fourier spectra, such that their power spectra 
cannot be described as a $\chi^{2}$ with 2 degrees of freedom.  
Therefore the statistics of the $2l+1$ power spectra of a given $l$ cannot be derived 
from the product of $2l+1$ $\chi^{2}$ with 2 degrees of freedom as in Appourchaux \etal (1995).  
Nevertheless, the real and imaginary parts of the Fourier 
spectra will still be normally distributed; in other words, the 
Fourier spectra have a \emph{multi-normal distribution} defined by a 
covariance matrix.  This 
fact will be used to derive the statistics of the observation.  The 
covariance matrix is the sum of the noise and mode covariance 
matrices, which are not necessarily the same.  Last but not least, 
the theoretical probability distribution has to be 
generated using the previous covariance matrix.

In summary, to understand the statistics of resolved observation, one has 
to follow four steps:
\begin{itemize}
\item Compute leakage matrices,
\item Compute mode covariance matrices (related to the leakage),
\item Compute noise covariance matrices,
\item Generate the likelihood from the theoretical probability distribution
\end{itemize}
Each step is described in detail hereafter.

\subsubsection{Leakage matrices}
Due to the spherical symmetry of the Sun, the most likely 
weights to be used to isolate the modes are the spherical harmonics $Y_{l,m}$.  Here we 
generalize the approach to any weight $W_{l,m}$.  The result is the 
observation of Fourier spectra $y_{m}^{l}(\nu)$ that are related to what we 
want to detect, i.e. the Fourier spectra of the individual Fourier 
spectrum $x_{m'}^{l'}(\nu)$, by the so-called leakage matrix (Schou 1992, Schou and Brown 1994).  The following expression can be derived for as many different 
degrees as needed; for simplicity we wrote it for 2 different degrees 
$l,l'$ as:
 \begin{equation}		
 	\vec{y}=\tens{\cal{C}}^{(l,l')}\vec{x}
 	\label{dbef}
 \end{equation}
where $\vec{x}(\nu)$ and $\vec{y}(\nu)$ are 2 complex vectors made 
each of $2l+2l'+2$ component: $2l+1$ 
components for $l$, $2l'+1$ components for $l'$ and 
$\tens{\cal{C}}^{(l,l')}$ is the leakage matrix of both $l$ and 
$l'$.  The dimension of $\tens{\cal{C}}^{(l,l')}$ is $(2l+2l'+2) 
\times (2l+2l'+2)$. The 
coefficient of the leakage matrix can be computed as:
 \begin{equation}		
 	\tens{\cal{C}}^{(l,l')}_{m,m'}=\frac{b^{(l,l')}_{m,m'}}{b^{(l',l')}_{m',m'}}
 	\label{d}
 \end{equation}
with:
  \begin{eqnarray}		
 	b^{(l,l')}_{m,m'}=\int_{\cal{D}}\frac{W_{l,m}^{*}(\theta,\phi)}{n_{l,m}} 
 	\tilde{Y}_{l',m'}(\theta,\phi)A(\theta,\phi) \sin \theta d\theta d\phi
 	\label{d6}
 \end{eqnarray}
where $m=-l,\ldots,l$, $m'=-l',\ldots,l'$, $*$ denotes the complex conjugate, $\theta,\phi$ are the angles 
in a spherical coordinate system, $\cal{D}$ is the integration domain, 
$\tilde{Y}_{l',m'}$ is the generalized velocity or intensity perturbation 
of the mode $(l',m')$, $A(\theta, \phi)$ 
is an apodization function, $n_{l,m}$ is a sensitivity correction factor 
associated with $W_{l,m}$.  The ratio ensures that $\tens{\cal{C}}^{(l,l)}_{m,m}=1$.
The apodization function $A$ is the product of 3 different function as:
\begin{equation}
A(\theta,\phi)=A_{n}(\theta,\phi)A_{d}(\theta,\phi)A_{a}(\theta,\phi)
\label{d7}
\end{equation}
$A_{n}$ is the natural apodization function due to the way the images are 
obtained: for intensity this is the limb darkening ($I(\mu)$), and 
for velocity the projection factor ($\mu=\sin \theta \cos \phi$).  
$A_{d}$ is the data analysis apodization: for data re-mapped on the 
Sun'surface it is unity; for 
no re-mapping, it is the projection factor ($\mu=\sin \theta \cos \phi$).
$A_{a}$ is the artificial apodization that can take into account the non-linear 
velocity (or intensity) response of the instrument over the solar disk, or can help to 
reduce limb effects.  Here we must point out that the leakage matrix has 
a useful property such as:
\begin{eqnarray}			
 	\tens{\cal{C}}^{(l,l')}_{m,m'}=\tens{\cal{C}}^{(l',l)*}_{m',m}\frac{b^{(l,l)*}_{m,m}}{b^{(l',l')}_{m',m'}}
 	\frac{b^{(l,l')}_{m,m'}}{b^{(l',l)*}_{m',m}}
 	\label{d1}
 \end{eqnarray}
It shows that $\tens{\cal{C}}^{(l,l')}$ is in general \emph{not} hermitian nor symmetrical.  
Nevertheless, when $W_{l,m}=\tilde{Y}_{l,m}$, it is possible with a 
proper sensitivity factor correction of $W_{l,m}$ to have 
such a property.  In this case the sensitivity correction is given by:
\begin{equation}
 	n_{l,m}=\sqrt{\int_{\cal{D}}\tilde{Y}_{l,m}^{*}(\theta,\phi) 
 	\tilde{Y}_{l,m}(\theta,\phi)A(\theta,\phi) \sin \theta d\theta d\phi}
\end{equation}
which is the `natural' normalization factor of the perturbation 
$\tilde{Y}_{l,m}$.  Of course in this latter case, we have: 
\begin{eqnarray}		
 	\tens{\cal{C}}^{(l,l')}_{m,m'}=\tens{\cal{C}}^{(l',l)*}_{m',m}
 	\label{dd1}
 \end{eqnarray}
Unfortunately, the leakage matrix does not always have 
such a nice property, especially because $W_{l,m}\ne\tilde{Y}_{l,m}$.  
This was the case for the ground-based Luminosity Oscillations Imager 
(LOI) (Appourchaux \etal, 1994) and for the GONG instrument 
(Hill, 1997, private communication).  In both cases, this is \emph{not} 
produced by the observation techniques but by the data analysis techniques.

If the 
weight functions $W_{l,m}$ and the observed perturbations $\tilde{Y}_{l,m}$ have the 
same symmetry properties as the spherical harmonics $Y_{l,m}$ (or if 
$W_{l,m}=\tilde{Y}_{l,m}=Y_{l,m}$), the leakage 
matrix is real as shown by Schou (1992).  In addition the leakage 
elements of $\tens{\cal{C}}_{m,m'}^{(l,l')}$ are zero if $l+m+l'+m'$ is 
odd; this is the case when the Sun is \emph{not} tilted with respect 
to the observer's axis of reference ($P$=0, $B$=0).  If the axes of reference 
of $W_{l,m}$ differ 
from that of the $Y_{l,m}$ these 2 properties can be lost.  For 
instance, an incorrect orientation of the Sun axis with respect to 
the detector axis could lead to a complex leakage matrix; or a Sun 
seen at an angle $B \neq 0$ give a real leakage matrix with non-zero elements with $l+m+l'+m'$ 
odd.  This latter property has been used by Gizon \etal 1997 to infer the inclination of 
the Sun's core.

Equation (\ref {d}) is valid when the size of the pixel is small compared with the 
spatial scale of the degree.  When the pixels are larger, one should 
write the following:
 \begin{equation}		
 	\tens{\cal{C}}^{(l,l')}_{m,m'}=\frac{n_{l',m'}}{n_{l,m}}
 	\frac{\sum_{i}w_{i}^{(l,m)*}\tilde{y}_{i}^{(l',m')} }
 	{\sum_{i}w_{i}^{(l',m')*}\tilde{y}_{i}^{(l',m')}}
 	\label{ds}
 \end{equation}
where the $\tilde{y}_{i}$ are given by:
 \begin{equation}		
 	\tilde{y}_{i}^{(l',m')}=\int_{{\cal{D}}_{i}}\tilde{Y}_{l'}^{m'}(\theta,\phi)
 	A(\theta,\phi) \sin \theta d\theta d\phi
 	\label{ddsmall}
 \end{equation}
where ${\cal{D}}_{i}$ is the area defined by the $i$-th pixel and 
$w_{i}^{(l,m)}$ is the weight applied to the $i$-th pixel to extract 
the $l,m$ mode.  Equation 
(\ref {ds}) is the more general form used for the LOI 
(Appourchaux and Andersen, 1990).  As a starting point, the $w_{i}^{(l,m)}$ can also be 
taken as the $\tilde{y}_{i}^{(l,m)}$.

\subsubsection{p-mode covariance matrix}
To compute the covariance of the complex vector $\vec{y}(\nu)$ as a real number we form 
 the vector $\vec{z}_{\vec{y}}(\nu)$ defined as:
 \begin{eqnarray*} 
 \vec{z}_{\vec{y}}^{\rm T}(\nu)=(\mbox{Re}(\vec{y}^{\rm T}),\mbox{Im}(\vec{y}^{\rm T}))
 \end{eqnarray*}
 In absence of noise, 
 the covariance matrix $\tens{M}(\nu)$ of the vector $\vec{z}_{\vec{y}}(\nu)$ can be generated using a 
 complex notation:
  \begin{equation}
\tens{M}^{(l,l')}(\nu)=\left(\begin{array}{cc}
\tens{\cal{M}}_{r}(\nu) & \tens{\cal{M}}_{i}(\nu)\\
-\tens{\cal{M}}_{i}(\nu) & \tens{\cal{M}}_{r}(\nu)\\
\end{array}\right).
\end{equation}
$\tens{M}^{(l,l')}$ is a super matrix where $\tens{\cal{M}}_{r}(\nu)$ and $\tens{\cal{M}}_{i}(\nu)$ are the real and imaginary 
parts of a complex matrix $	\tens{\cal{M}}^{(l,l')}$ which elements are given by:
 \begin{equation}	 	 	
 	\tens{\cal{M}}^{(l,l')}_{m,m'}(\nu)=\sum_{l''=l,l'}\sum_{m''=-l''}^{l''}\tens{\cal{C}}^{(l'',l')}_{m'',m'}
 	\tens{\cal{C}}^{(l'',l)*}_{m'',m}f_{m''}^{l''}(\nu)
 	\label{g}
 \end{equation} 
where $f_{m''}^{l''}(\nu)$ is the variance of the $l'',m''$ mode which 
profile is given by 
Eq. (\ref{c}), in which $\nu_{0}$ is a function of $m$.  The real and 
imaginary parts of Eq. (\ref{g}) 
will give respectively the covariance of the 
 real (or imaginary) part of $\vec{y}$, and the covariance between
  the real and imaginary part of $\vec{y}$.  It is 
 obvious from Eq. (\ref{g}) that $\tens{\cal{M}}^{(l,l')}$ is hermitian.

Schou (1992) 
gave an equation similar to Eq. (\ref{g}) for a real leakage matrix 
and for a single degree.  Here we add a subtlety 
to the formulation of Schou (1992), the matrix $\tens{M}^{(l,l')}(\nu)$ can be 
decomposed as follows:
 \begin{equation}	
 	\tens{M}^{(l,l')}(\nu)=\left(\begin{array}{cc}
\tens{v}(\nu) & \tens{w}(\nu)\\
-\tens{w}(\nu) & \tens{v}(\nu)\\
\end{array}\right)\left(\begin{array}{cc}
\tens{v}^{\rm T}(\nu) & -\tens{w}^{\rm T}(\nu)\\
\tens{w}^{\rm T}(\nu) & \tens{v}^{\rm T}(\nu)\\
\end{array}\right)
 	\label{h}
 \end{equation} 
where $\rm T$ is the transpose of a matrix.  The elements of $\tens{v}$ 
and $\tens{w}$ are given by:
 \begin{equation}	
 	\tens{v}^{(l,l')}_{m,m'}(\nu)=\sqrt{f_{m}^{l}(\nu)}\mbox{Re}(\tens{\cal{C}}^{(l,l')}_{m,m'})
 	\label{i}
 \end{equation}
 \begin{equation}	
 	\tens{w}^{(l,l')}_{m,m'}(\nu)=\sqrt{f_{m}^{l}(\nu)}\mbox{Im}(\tens{\cal{C}}^{(l,l')}_{m,m'})
 	\label{ii}
 \end{equation}
We will see later on that this decomposition is of prime importance 
for  understanding the statistics of the observation.

\subsubsection{Noise covariance matrix}
Unfortunately, the observed vector $\vec{y}(\nu)$ include a noise 
contribution.  Due to the way the data are combined, the noises between the 
different $2l+2l'+2$ components of this vector are also correlated.  Schou (1992) gave the
correlation matrix when the filter used are spherical harmonics $Y_{l,m}$.  
A more general formulation can be written as:
 \begin{equation}
\tens{B}^{(l,l')}(\nu)=\left(\begin{array}{cc}
\tens{\cal{B}}_{r}(\nu) & \tens{\cal{B}}_{i}(\nu)\\
-\tens{\cal{B}}_{i}(\nu) & \tens{\cal{B}}_{r}(\nu)\\
\end{array}\right)
\label{dddd}
\end{equation}
$\tens{B}^{(l,l')}$ is a super matrix where $\tens{\cal{B}}_{r}(\nu)$ and $\tens{\cal{B}}_{i}(\nu)$ are the 
real and imaginary parts of the complex matrix 
$\tens{\cal{B}}^{(l,l')}$.  The dimension of $\tens{\cal{B}}^{(l,l')}$ is 
$(2l+2l'+2) \times (2l+2l'+2)$.  Its elements are given by:
 \begin{eqnarray}		
 	\tens{\cal{B}}^{(l,l')}_{m,m'}=\int_{\cal{D}}\frac{W_{l,m}^{*}(\theta,\phi)}{n_{l,m}} 
 	\frac{W_{l',m'}(\theta,\phi)}{n_{l',m'}}
 	a(\theta,\phi)  \sin \theta d\theta d\phi
 	\label{dd}
 \end{eqnarray}
with
 \begin{eqnarray}		 
 a(\theta,\phi)=A_{a}^{2}(\theta,\phi)A_{d}(\theta,\phi)\sigma^{2}_{\odot}(\theta,\phi,\nu)
 	\label{dd2}
 \end{eqnarray}
where $a$ is an apodization function which characterizes through 
$\sigma^{2}_{\odot}(\theta,\phi,\nu)$ how the noise varies over the solar 
image, assuming that the noise is uncorrelated between different 
points on the Sun; $A_{a}, A_{d}$ are defined in Eq. (\ref{d7}).  When the instrumental noise is low, $a$ 
is derived from the characteristics of the solar noise.  
The evaluation of $\tens{\cal{B}}^{(l,l')}$ is less 
straightforward than that of $\tens{\cal{C}}^{(l,l')}$ because we need to 
know a model of the solar noise.  An easier way to understand the noise correlation 
is to built the 
ratio covariance matrix or `pseudo' noise leakage matrix 
$\tens{\cal{R}}$ as:
 \begin{equation}			
 	\tens{\cal{R}}^{(l,l')}_{m,m'}=\frac{\tens{\cal{B}}^{(l,l')}_{m,m'}}{\tens{\cal{B}}^{(l',l')}_{m',m'}}
 	\label{dn}
 \end{equation}	
Here we can see the similarity between $\tens{\cal{R}}$ and $\tens{\cal{C}}$.  In velocity, 
the granulation noise is rather low at the center of the 
disk and then increases towards the limb; the meso- and 
super-granulation exhibits somewhat different or complementary 
center-to-limb variations.  In intensity, the 
granulation noise is a function of the number of granules; the noise 
is larger at the center of the disk and decreases slowly towards the 
limb.  In addition the solar noise in intensity has no contribution 
from mesogranulation (Fr\"ohlich \etal, 1997), making the spatial dependence of the 
noise almost independent of frequency across the p-mode range.  This is not the case in velocity 
where mesogranulation still contributes to the noise in the p-mode 
range.  Therefore in intensity the apodization $a$ is closer to $A$ than 
in velocity, making the ratio covariance matrix $\tens{\cal{R}}^{(l,l')}$ very close to the 
leakage matrix $\tens{\cal{C}}^{(l,l')}$.  Although $\tens{\cal{R}}^{(l,l')}$ 
is not mathematically useful, it is a matrix easy to visualize and 
understand (See Part II).  The ratio matrix has some properties of the 
leakage matrix like being not necessarily hermitian.  This is not the 
case of $\tens{\cal{B}}^{(l,l')}$ which is hermitian by definition.

Again, when the size of the pixel is large compared with the 
spatial scale of the degree, Eq. (\ref {dd}) is rewritten as 
follows:
 \begin{equation}		
 	\tens{\cal{B}}^{(l,l')}_{m,m'}(\nu)=\sum_{i}w_{i}^{*(l,m)}w_{i}^{(l',m')}b_{i}(\nu) 
 	\label{nds}
 \end{equation}
where $b_{i}$ is the variance of the noise of pixel $i$.  Equation 
(\ref {nds}) is the more general form used for the LOI.   

\subsubsection{Probability density of the observation and likelihood}
The statistical distribution of the Fourier spectra or of the
vector $\vec{z}_{\vec{y}}$ is a multi-normal distribution.  The 
probability density is given by:
 \begin{equation}	  	  	
  	p_{\vec{y}}(\nu)=\frac{e^{-\frac{1}{2}\vec{z}_{\vec{y}}^{T}(\nu)\tens{V}^{-1}(\nu)
  	\vec{z}_{\vec{y}}(\nu)}}{(2\pi)^{d/2}\sqrt{|\tens{V}(\nu)|}}
  	\label{fp}
  	 \end{equation} 
where $d$ is the number of elements of $\vec{z}_{\vec{y}}$, $\tens{V}$ is a short notation for the following matrix: $\tens{V}^{(l,l')}(\nu)=
\tens{M}^{(l,l')}(\nu)+\tens{B}^{(l,l')}(\nu)$; this is the matrix given 
by the sum of the p-mode and noise covariance matrix; the p modes and 
the noises are assumed to be independent of each other.  The matrix 
$\tens{V}^{(l,l')}(\nu)$ can also be built from sub-matrices as: 
$\tens{\cal{V}}^{(l,l')}=\tens{\cal{M}}^{(l,l')}+\tens{\cal{B}}^{(l,l')}$; 
as a result $\tens{\cal{V}}^{(l,l')}$ is also hermitian.  Equation 
(\ref{fp}) is the most general formulation for any multi-normal 
distribution with a given covariance matrix $\tens{V}$ (Kendall and 
Stuart, 1967).

Using Eq. (\ref{fp}), we can write the likelihood $L$ of an observation
 of $\vec{z}_{\vec{y}}(\nu_{i})$ at $N$ different frequencies 
$\nu_{i}$ as given by:
 \begin{equation}	  	  	
  	L_{\vec{y}}^{(l,l')}=\prod_{i=1}^{N}\frac{e^{-\frac{1}{2}\vec{z}_{\vec{y}}^{T}(\nu_{i})\tens{V}^{-1}(\nu_{i})
  	\vec{z}_{\vec{y}}(\nu_{i})}}{(2\pi)^{d/2}\sqrt{|\tens{V}(\nu_{i})|}}
  	\label{f}
  	 \end{equation} 
We assumed that the frequency bins are independent of each other.  
This is the case when the data have no gaps.  For unresolved 
observation having gap, 
the expression of the likelihood becomes extremely complicated as shown by 
Gabriel (1994) .  For resolved observation 
having gaps, as for the LOWL data of Tomczyk \etal (1995), it is impracticable 
to use the full formulation of the likelihood: Tomczyk \etal (1995) used 
Eq. (\ref{f}) as an approximation for fitting the LOWL data.

In principle, given the observed vector $\vec{y}$, it is always possible in 
the absence of noise to recover the vector $\vec{x}$.  Due to the presence 
of noise only a solution close to the ideal one can be found that 
will minimize the correlation between the components.  Provided that 
the leakage matrix can be inverted, we have by analogy to Eq. 
(\ref{dbef}):
 \begin{equation}	
\vec{\tilde{x}}=\tens{\cal{C}}^{-1}\vec{y}
 	\label{j2}
 \end{equation}
where $\tens{\cal{C}}=\tens{\cal{C}}^{(l,l')}$.  
Then we can write a similar equation for $\vec{z}_{\vec{y}}$ and 
$\vec{z}_{\vec{\tilde{x}}}$ as:
 \begin{equation}	
\vec{z}_{\vec{\tilde{x}}}=\tens{C}^{-1}\vec{z}_{\vec{y}}
 	\label{jj2}
 \end{equation}
where $\tens{C}$ is defined as:
 \begin{equation}
\tens{C}^{(l,l')}(\nu)=\left(\begin{array}{cc}
\tens{\cal{C}}_{r}(\nu) & \tens{\cal{C}}_{i}(\nu)\\
-\tens{\cal{C}}_{i}(\nu) & \tens{\cal{C}}_{r}(\nu)\\
\end{array}\right)
\label{dddd2}
\end{equation}
$\tens{C}^{(l,l')}$ is a super matrix where $\tens{\cal{C}}_{r}$ and $\tens{\cal{C}}_{i}$ are the 
real and imaginary parts of the complex matrix 
$\tens{\cal{C}}^{(l,l')}$.  Using Eq. 
(\ref{jj2}) 
to replace $\vec{z}_{\vec{y}}$ by $\vec{z}_{\vec{\tilde{x}}}$ in Eq. (\ref{f}) we can rewrite 
this latter as:
 \begin{equation}	  	
  	L_{\vec{y}}^{(l,l')}=\prod_{i=1}^{N}\frac{e^{-\frac{1}{2}\vec{z}_{\vec{\tilde{x}}}^{T}(\nu_{i})\tens{V'}^{-1}(\nu_{i})  	  	
   	  	\vec{z}_{\vec{\tilde{x}}}(\nu_{i})}}{(2\pi)^{d/2}\sqrt{|\tens{V'}(\nu_{i})|}}
   	  	\frac{1}{|\tens{C}|}=\frac{1}{|\tens{C}|^{N}}L_{\vec{\tilde{x}}}^{(l,l')}
 	\label{j}
 \end{equation}
with $\tens{V'}$ given by:
\begin{equation}
\tens{V'}=\tens{C}^{-1} \tens{V} {\tens{C}^{\rm T}}^{-1}=
\tens{C}^{-1} \tens{M}^{(l,l')} {\tens{C}^{\rm T}}^{-1}+\tens{C}^{-1} 
\tens{B}^{(l,l')} {\tens{C}^{\rm T}}^{-1}
\label{linear}
\end{equation}
We recognize in Eq (\ref {j}) the probability density of the vector 
$\vec{z}_{\vec{\tilde{x}}}(\nu)$ to a 
constant (i.e. $|\tens{C}|^{-N}$).  As a matter of fact, it is well 
known that using a linear transformation similar to that of Eq. 
(\ref{jj2}) will produce the new covariance matrix $\tens{V'}$ of 
$\vec{z}_{\vec{\tilde{x}}}$ as written in Eq (\ref{linear}) (Davenport and 
Root, 1958).  It can be easily shown using Eqs. (\ref{h}) and (\ref{linear}) that the matrix 
$\tens{D}(\nu)=\tens{C}^{-1} \tens{M}^{(l,l')} {\tens{C}^{\rm T}}^{-1}$ is 
diagonal and its element are given by:
 \begin{equation}	
  \tens{D}_{m'',m''}(\nu)=f_{m''}^{l''}(\nu)
  \label{k}
 \end{equation}
where $l''=l$ or $l'$ and $m''=-l'',\ldots,l''$.  Therefore Eq. (\ref {linear}) is the sum of a diagonal matrix representing
 the correlation between the p modes; and of a new noise 
covariance matrix representing the correlation of the 
components of the vector $\vec{\tilde{x}}$ after the transformation of 
Eq. (\ref {j2}).  It means that $\vec{\tilde{x}}$ has no 
correlation due to the p modes as we could expect from Eq. (\ref{j2}): 
the leakage matrix of $\vec{\tilde{x}}$ is the identity matrix.  
In summary, \emph{there is no gain in 
fitting data for which the leakage matrix is the identity matrix: the 2 approaches are 
identical}.  The main problem is really to know the leakage matrices, 
not only theoretically but also experimentally: this is the subject of 
the Part II.   

It can be derived from Eq. (\ref{linear}) that it is also possible to 
remove correlation due to the noise by replacing $\tens{C}$ by a 
proper matrix associated with $\tens{B}^{(l,l')}$.  The derivation of 
this matrix is given in Appendix A.

\subsubsection{The use of the likelihood in practice}
When a single degree is observed, it is quite simple to maximize the 
likelihood of Eq. \ref{f} using $\vec{y}$, or using $\vec{\tilde{x}}$ 
as in Eq. \ref{j}.  For low degree and low frequency modes, this is possible for $l=0, 
2, 3$.  As soon as the mode linewidth increases, at high 
frequencies, the assumption of a single degree is not valid anymore.  
For example, $l=0$ and $l=1$ overlap with $l=2$ and $l=3$, respectively.
At high frequencies, the effect of the aliasing degree should be 
taken into account.

For the other low degree modes, the likelihood 
becomes somewhat more complicated.  It is well known, that in the $(m,\nu)$ diagramme of $l=1$, there are 
leaks coming from $l=6$ and $l=9$; in the $(m,\nu)$ diagramme of $l=4$, there are 
leaks of $l=7$ and vice versa (Appourchaux \etal, 1997).  The leaks 
have severe effects on determination of the p-mode parameters of the $l=1$. 

When many degrees are overlapping, one should use Eq. (\ref{f}) using the covariance matrix 
for $l$ and $l'$.  Nevertheless, we do not 
advice to do so for fitting the p 
modes; it has some severe computer speed penalty.  Instead we advice to clean the data 
by inverting the full leakage matrix taking into account the effects 
of the various degrees on each other, in a similar way to Eq. 
(\ref {j2}).  This technique has been applied to the LOI and GONG data, and is 
developed in Part II. 

Last but not least, when the signal-to-noise 
ratio is high (i.e. we neglect $\tens{B}^{(l,l')}$ in Eq. 
\ref{linear}), the elements of the vector $\vec{z}_{\vec{\tilde{x}}}(\nu)$ 
are all independent of each other, leading to a statistical distribution which is 
a product of $\chi^{2}$ with 2 degree of freedom.  This is an 
approximation which is useful and less incorrect that using this 
statistics for the GONG data for the vector $\vec{z}_{\vec{y}}(\nu)$ as in Hill \etal 
1996.

\section{Monte-Carlo simulations}
\subsection{Why are they needed?}
Before applying Eq. (\ref{f}) to real data, it is always advisable 
to test the power of MLE on synthetic data, i.e. performing 
Monte-Carlo simulations.  They are not merely for playing games; these 
simulations are real tools for understanding what we fit and how we fit it.
Assuming that the statistics of the real solar spectra is known, performing Monte-Carlo 
is useful for the following reasons:
\begin{itemize}
\item Assessing the model of the mode and noise covariance
\item Assessing the statistical distribution of the parameters
\item Assessing the precision of parameters
\end{itemize}
First, the model of the covariances can be imperfect.  The effect of 
an imperfect knowledge of the covariance can help us understand how 
these will influence the determination of the parameters, i.e. deriving the sensitivity of the systematic errors to this imperfect 
knowledge.  
Second, the parameters derived by the MLE should have the desirable 
properties of having a normal distribution; if not we advise to apply 
a change of fitted parameters.  For example, as we will see later on, 
we do not fit the linewidth itself but the log of the linewidth.  A 
normal distribution is necessary to derive meaningful error bars, 
this is the assumption behind Eq. (\ref{hij}).
Third, in order to be able to derive a good estimate of the error 
bars using one realization, the standard deviation of a large sample of fitted 
parameters should be equal to the mean of formal errors return by the 
fit (See Eq. (\ref {large})).

\begin{figure*}[!]
\centerline{\hbox{
\psfig{file=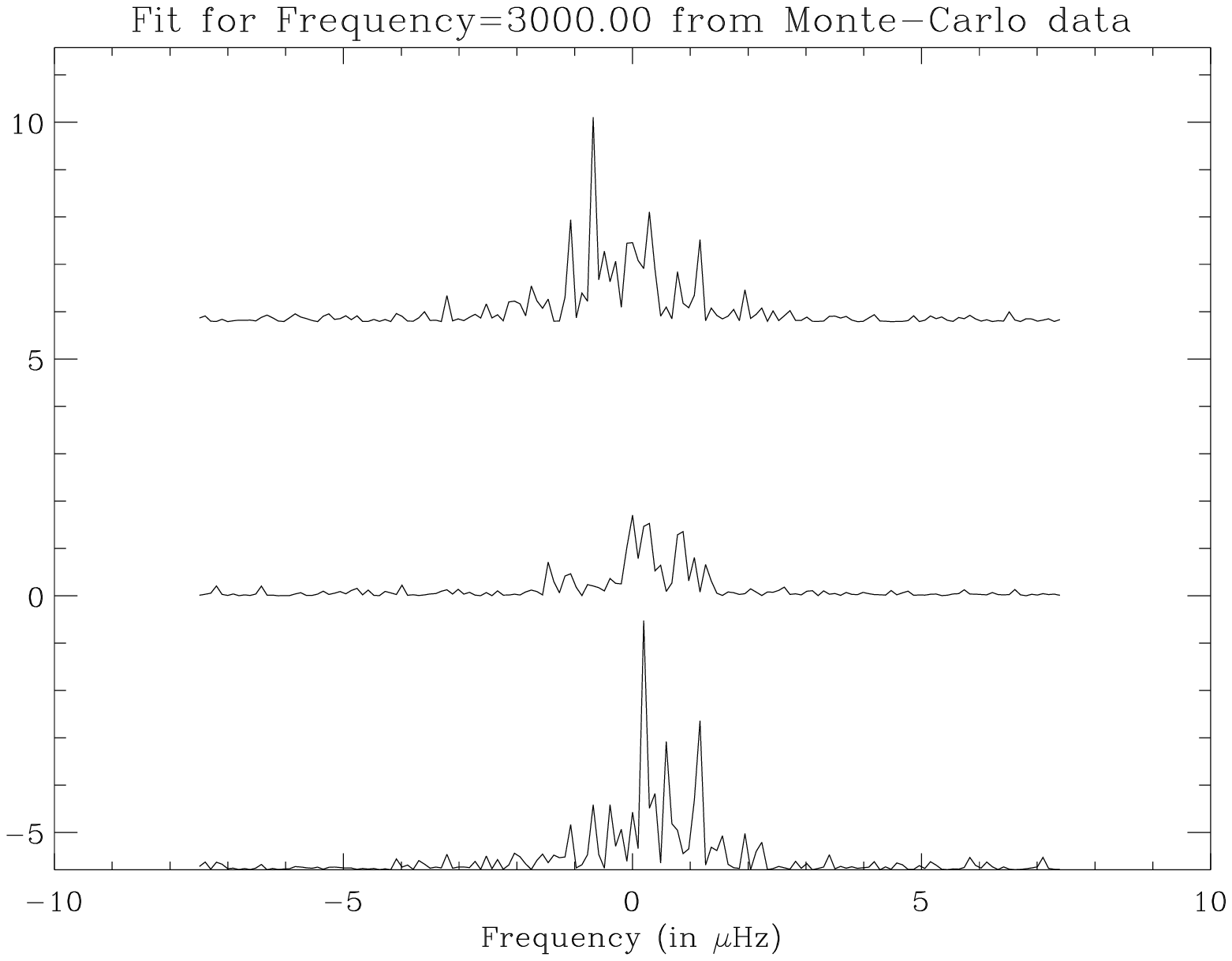,width=8.0cm}
\psfig{file=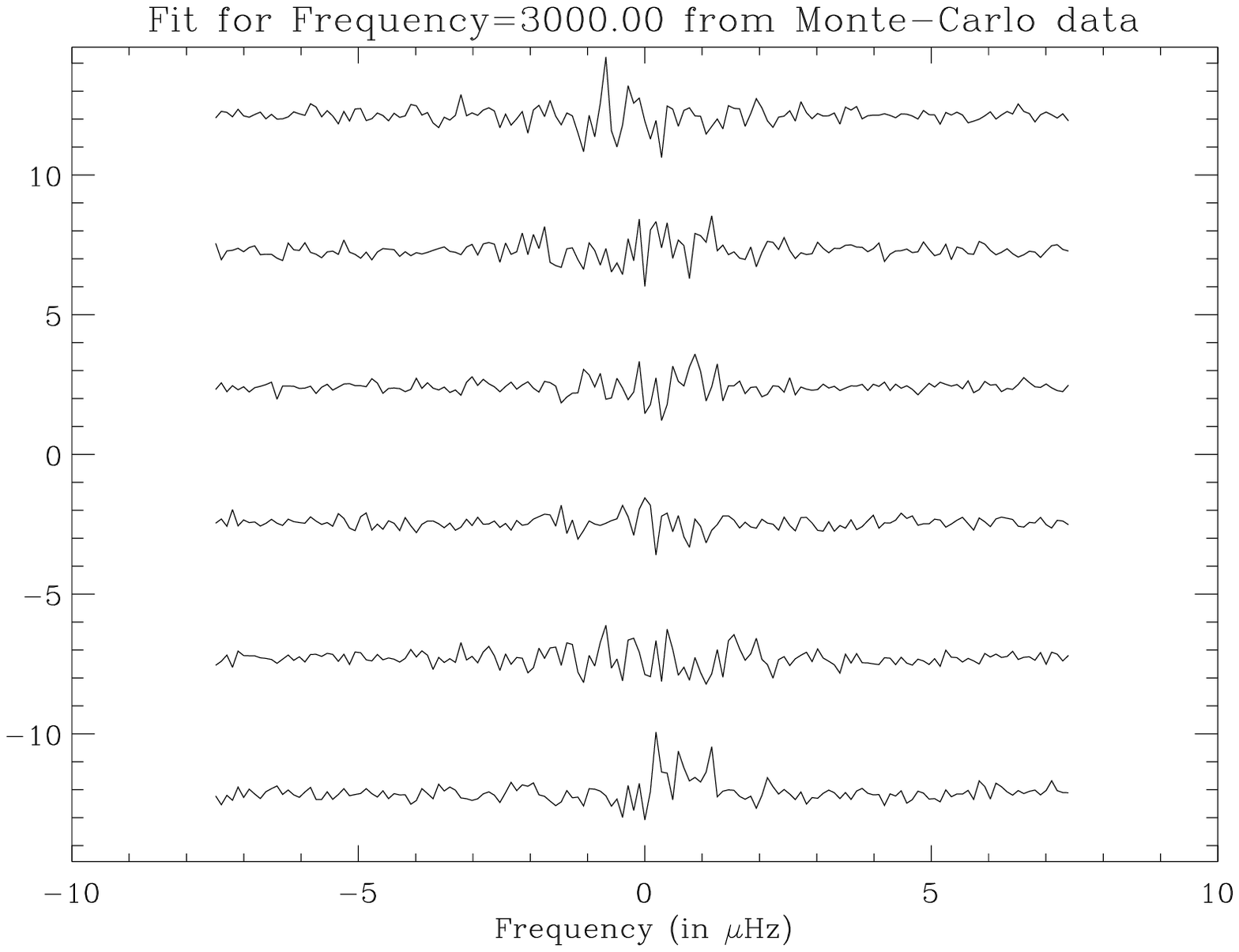,width=8.0cm}
}} 
\caption{(Left) Power spectra of a synthetic $l=1$ as it would be 
observed by the LOI.  The frequency resolution corresponds to 4 months 
of data.  The signal-to-noise ratio is about 20-30. The traces from 
bottom to top corresponds to m=-1,0,+1. (Right) Fourier spectra for 
$l=1$ (same data). The first, third and fifth traces from the bottom 
represents the real part of the spectrum of $m=-1,0$ and 1, 
respectively; the other traces are the imaginary parts.  The leakage 
between $m=-1$ and $m=+1$ is 0.45 in the Fourier spectra.}
\label{spectra} 
\end{figure*}

\subsection{Generation of synthetic data for the LOI}
The performance of this instrument has been described in Appourchaux et 
al (1997).  Briefly, it is a small instrument made of 12 pixels for 
detecting solar intensity fluctuations.  The p-mode signals were 
generated in the Fourier spectra by using the following:
  \begin{equation}	  	
  	\vec{y}(\nu)=\tens{\cal{C}}^{(l,l)}\vec{x}(\nu)+\sum_{i=1}^{N_{pix}}\vec{\tilde{y}}_{i}^{l}p_{i}
 	\label{jz}
  \end{equation}
where $\vec{y}$ is the observed vector of $2l+1$ Fourier spectra, 
$\tens{\cal{C}}^{(l,l)}$ is the leakage matrix given by Eq. 
(\ref{ds}), $\vec{x}$ is a complex random vector with $2l+1$ components (each 
component represents the signal of an $l,m$ mode, with uncorrelated real and imaginary 
part), $\vec{\tilde{y}}_{i}^{l}$ are computed as in Eq. (\ref{ddsmall}) 
using spherical harmonics, and $p_{i}$ is the 
noise for a given pixel $i$.  The variance of the real or imaginary part 
of the $m$-th 
component of $\vec{x}$ is given by $f_{m}^{l}(\nu)$; the mean of $\vec{x}$ is 0.  The function 
$f_{m}^{l}(\nu)$ 
describes the profile of each $m$ which is displaced from $m$ by an 
amount which is given by:
  \begin{equation}	
  	\nu_{m}=\nu_{0}+l\sum_{i=1}^{5}a_{i}{\cal{P}}_{i}^{(l)}(m/l)
 	\label{jj1}
 	\end{equation}
 where the ${\cal{P}}_{i}^{(l)}$ are derived from the Clebsch-Gordan 
 coefficients, the expression of which can be found in Ritzwoller and 
 Lavelly (1991); they are normalized such that  
 ${\cal{P}}_{i}^{(l)}(1)=1$.  Here we assumed a common linewidth for the $l,n$ mode, 
 and different amplitudes for the $2l+1$ components.
 The profile are symmetrical in the shape of a lorentzian.
 
 The variance of the pixel noise is assumed to be the same for the 
 pixels with the same shape.  The mean of the pixel noise is 0.  
 For the LOI with its 12 pixels, there are 3 different shapes giving 3 
 independent noises.   
 
 After generating the synthetic signals according to Eq. 
 (\ref{jz}), the data are fitted by minimizing the likelihood of 
 Eq. (\ref{f}).  Figure \ref{spectra} shows an example of Fourier 
 spectra generated synthetically.  The typical signal-to-noise ratio 
 in the power spectra is about 20-30.  
The frequency resolution is equivalent to 4 months of data.  We 
performed 1000 simulations of the spectra.

\subsection{Results}
\subsubsection{For the nominal leakage matrix}
\begin{figure*}[!]
\centerline{\psfig{file=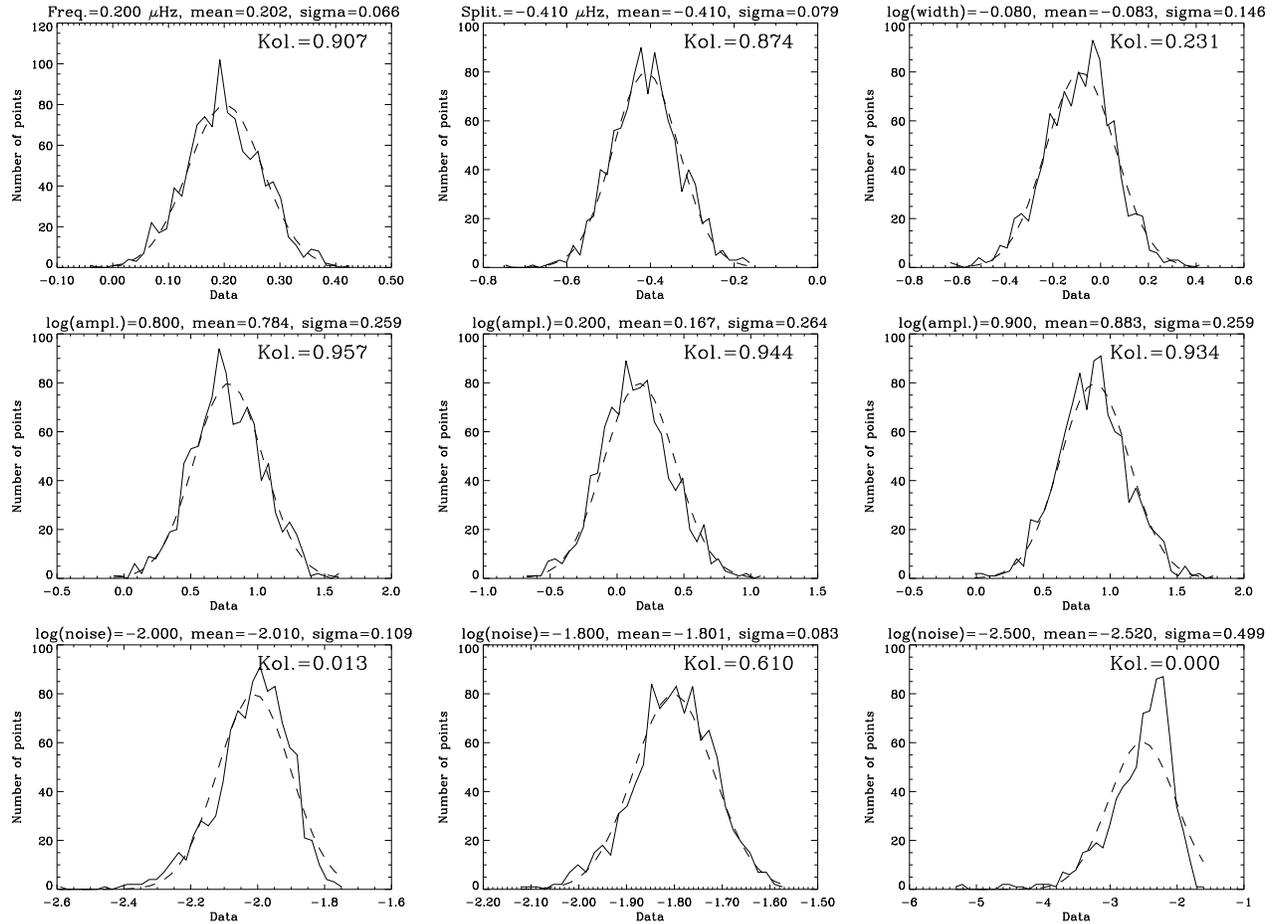,angle=90,width=17cm}}
\caption{Histograms for the fitted parameters:  (Plain line) Data, 
(Dashed line) Normal distribution with the same mean and $\sigma$ as 
the fitted parameters. (Top) Frequency (in $\mu$Hz), 
splitting $a_{1}$ (in $\mu$Hz), $\log(\gamma)$ ($\gamma$ in 
$\mu$Hz); (Middle) $\log$(Amplitude) for 
m=-1,0,1; (Bottom) $\log$(pixel noise).  For each histogram, the 
target value, the mean fitted value and the 1-$\sigma$ fitted valued 
are displayed.  The Kolmogorov-Smirnov test (Kol.) is displayed for each 
histogram; a number close to 0 show that the distribution is not normal.}
\label{mean} 
\end{figure*}

\begin{figure*}[!]
\centerline{\psfig{file=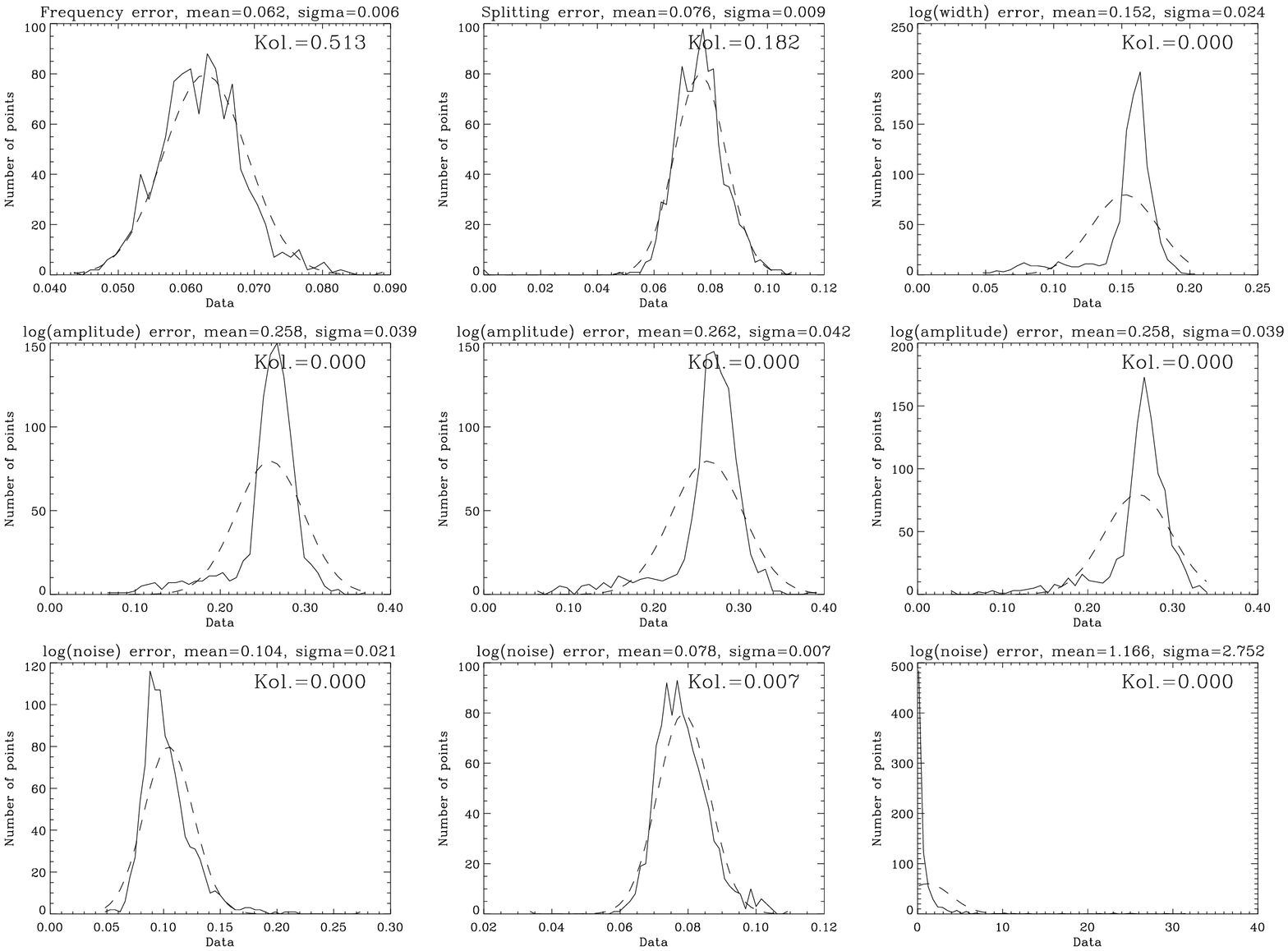,angle=90,width=17cm}}
\caption{Histograms for the error bars: (Plain line) Data, 
(Dashed line) Normal distribution with the same mean and $\sigma$. (Top) 
Frequency error (in $\mu$Hz), 
splitting $a_{1}$ error (in $\mu$Hz), $\log(\gamma)$ error ($\gamma$ in $\mu$Hz); (Middle) 
$\log$(Amplitude) 
error for 
$m$=-1,0,1; (Bottom) $\log$(pixel noise) error.  For each histogram, the 
target value, the mean fitted value and the 1-$\sigma$ fitted valued 
are displayed.  The Kolmogorov-Smirnov test is displayed for each 
histogram; a number close to 0 show that the distribution is not normal.}
\label{sigma} 
\end{figure*}

The data are fitted assuming a perfect knowledge of the leakage and noise 
covariance matrices, i.e. we know what we fit.   Figure \ref{mean} 
shows the distribution of the fitted parameters: the central frequency, 
splitting, $\log$(linewidth), 3$\times \log$(amplitude), 
3$\times \log$(pixel noise).  For the last 7 parameters, we fit the $\log$ of the 
parameter because this transformation give a statistical distribution 
closer to a normal distribution (or log-normal distribution).  It can 
be observed that the parameters derived are in most cases unbiased.
Figure \ref{sigma} shows the distribution of the error bars returned 
by the fit.  In most case the mean of the error bars (returned by the 
fit) is not very 
different from the 1-$\sigma$ deviation of the parameter distribution.
Similar simulations have been performed for various degree (up $l=3$).
They show the same typical results as for Figs. \ref{mean} and 
\ref{sigma}, i.e. the fitted parameters are not, or weakly, biased, and 
the error bars returned by the fit give a good estimate of the real 
error bars.

\subsubsection{Influence of a wrong leakage matrix}
As was shown by Eq. (\ref{j}), fitting p-mode spectra for which 
the leakage matrix is explicitly diagonal is equivalent to fitting 
p-mode spectra for which the matrix is {\it not} diagonal.  Of 
course, it is always possible to construct data with a purely diagonal 
leakage matrix using Eq. (\ref{j2}), but we do so assuming that we 
know the leakage matrix $\tens{\cal{C}}$.  As a matter of fact, what 
matters is not to have the identity matrix as leakage matrix, but 
more the knowledge of the latter.  

\begin{figure*}[!]
\centerline{\psfig{file=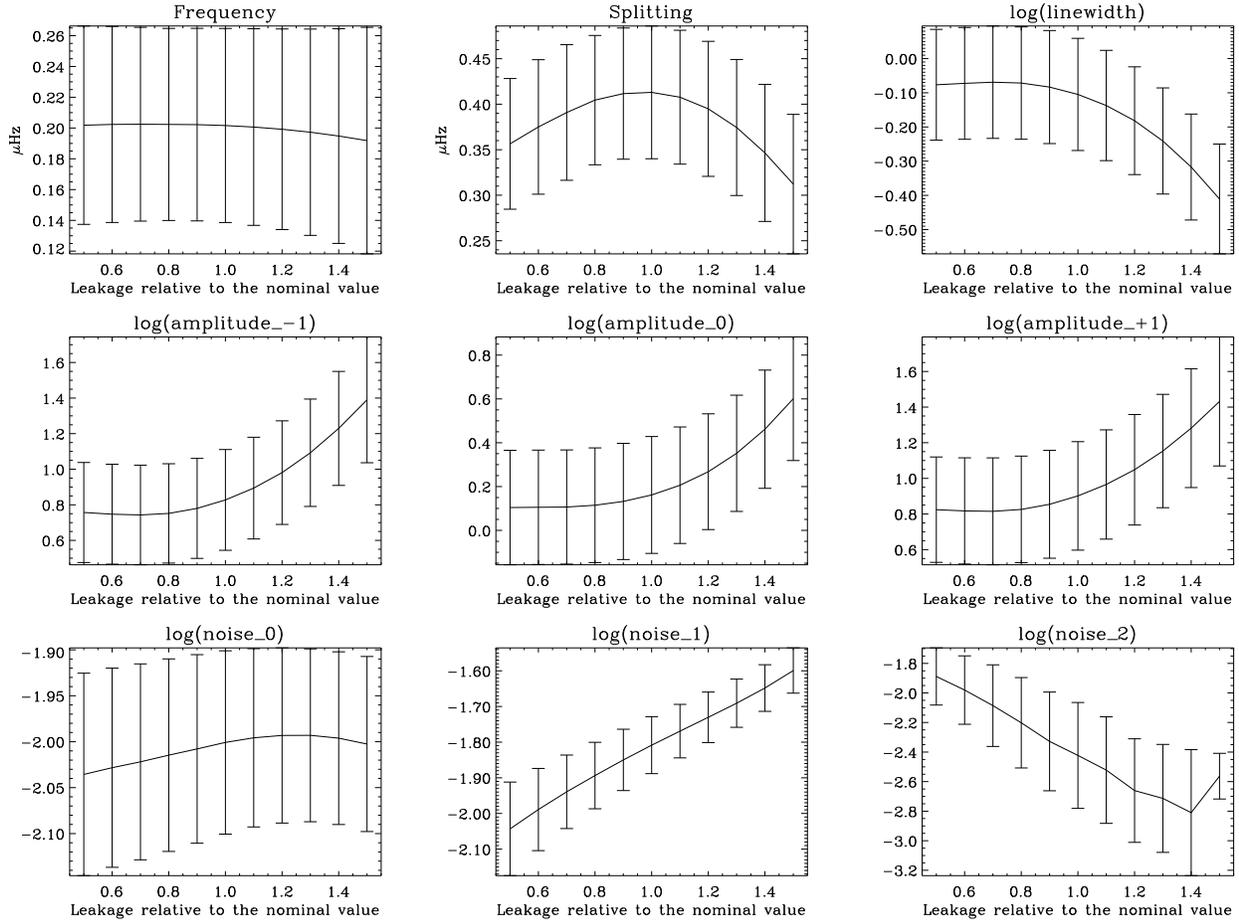,angle=90,width=17cm}}
\caption{Influence of the fitted parameters to relative changes of the assumed 
leakage element between $m=-1$ and $m=+1$ for $l=1$. (Top) Frequency, 
splitting $a_{1}$, $\log(\gamma)$; (Middle) $\log$(Amplitude) for 
m=-1,0,1; (Bottom) $\log$(pixel noise).  The target parameters are the 
same as for Fig. \ref{mean}.  Please note the parabolic 
shape for the splitting.}
\label{sys} 
\end{figure*}

Hereafter, we have investigated the 
influence of a wrongly assumed leakage matrix on the fitted parameters 
of $l$=1.  We made 100 realizations and change the leakage parameter 
between $m=-1$ and $m=+1$ by $\pm 50$\% from a nominal value for the LOI of 0.45.
Figure \ref{sys} shows the influence of varying the assumed leakage 
element on the fitted parameters.  It is quite interesting to note 
that the inferred central frequency is insensitive to mistakes in the 
leakage matrix.  The linewidth becomes underestimated when the error 
is larger than 20 \%, while the amplitudes become overestimated.  The most 
important result is the fact that the systematic error made on the 
splitting is not linear but quadratic. This systematic error can become as 
large as the error bars.  For example, with 1 year of LOI data and 
averaging over 10 modes, the error bars on the mean splitting is about 
15 nHz; this should be compared to a systematic error of 10 nHz for an 
error of 10\% of the $l=1$ leakage elements.    

\begin{figure*}[!]
\centerline{\psfig{file=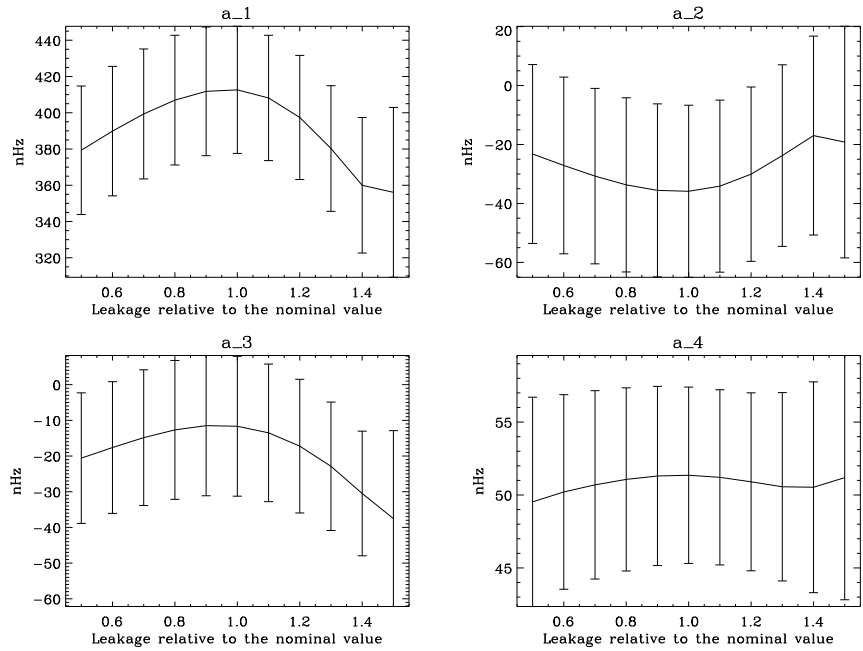,angle=90,width=12cm}}
\caption{Influence of the fitted splitting parameters to relative changes of the assumed 
of the assumed off-diagonal leakage element for $l=2$. (Top,left) 
$a_{1}$, target value: 410 nHz; (Top, right) $a_{2}$, target value: -30 
nHz; (Bottom, left) $a_{3}$, target value: -10 nHz; (Bottom, right) 
$a_{4}$, target value: +50 nHz.}
\label{sys2} 
\end{figure*}

Another test similar to that of the $l=1$ 
was performed with the $l=2$ mode.  We have assumed that 
all the off-diagonal elements of the leakage matrix were wrong by the 
same fixed amount.  Figure 
\ref{sys2} shows the results only for the splitting coefficients (from $a_{1}$ to 
$a_{4}$).  The other parameters 
linewidth, amplitudes and noises behave in the same manner as for 
$l=1$.  The systematic error on the splitting has also the same 
quadratic dependence as for $l=1$.  For $l=2$ the splitting error bars 
are typically $\sqrt{5}$ smaller than for $l=1$.  In this case the 
systematic errors become larger than the error bars, and therefore 
start to influence the inverted solar rotation.

It means that 
it is quite easy to underestimate the splitting whenever we under- or 
overestimate the leakage element. As a matter 
of fact, this behaviour was also found in the GONG data for $l=1$ and 
2 (Rabello-Soares and Appourchaux, 1998, in preparation).  
On the other hand, errors in the 
leakage matrix will not result in overestimating the splitting.  If 
the splitting is overestimated, the most likely source should be the 
presence of other degrees not taken into account in the analysis.  

\begin{figure*}[t]
\centerline{\psfig{file=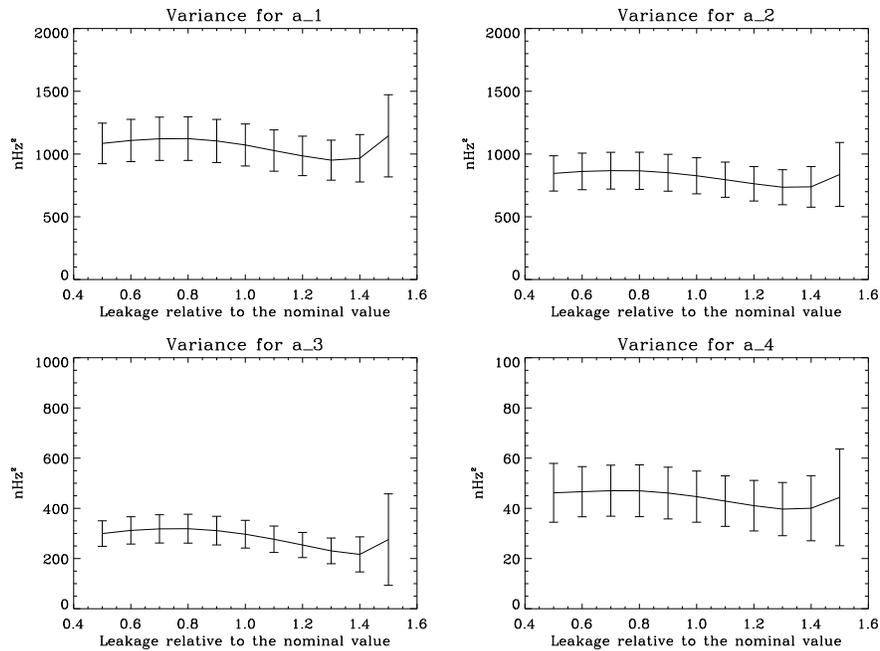,angle=90,width=12cm}}
\caption{Diagonal elements of the covariance matrix of the splitting 
coefficient, for $l=2$.  They are given as a function 
of the relative change of the assumed 
off-diagonal leakage element. (Top, left) For $a_{1}$; (Top, right) 
For $a_{2}$; (Bottom, left) For $a_{3}$; (Bottom, right) For $a_{4}$.}
\label{corr1} 
\end{figure*}

\begin{figure*}[!]
\centerline{\psfig{file=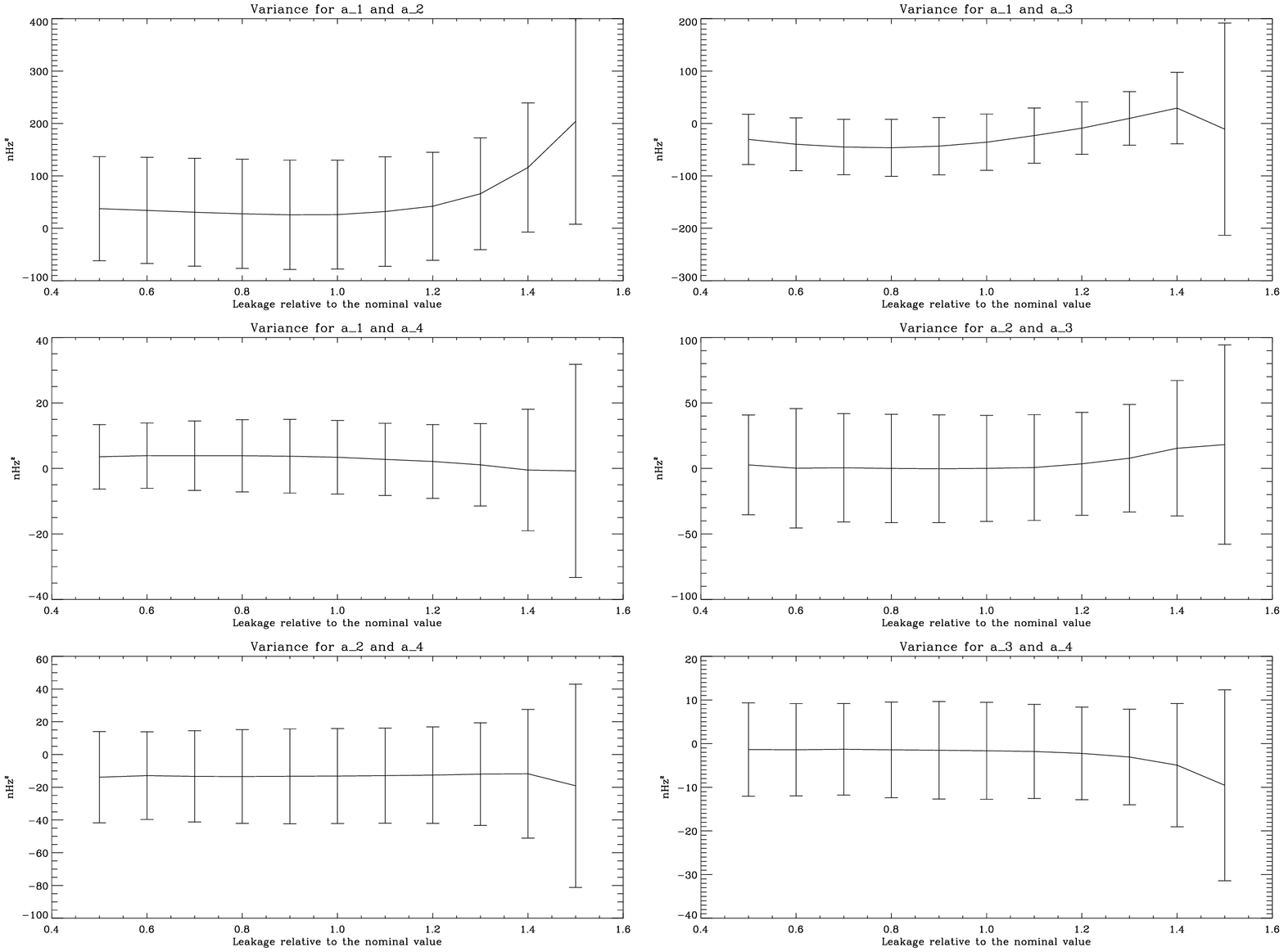,angle=90,width=17cm}}
\caption{Off-diagonal elements of the covariance matrix of the splitting 
coefficient, for $l=2$.  They are given as a function 
of the relative change of the assumed 
off-diagonal leakage element. (Top, left) For $a_{1}$ and $a_{2}$; (Top, right) 
For $a_{1}$ and $a_{3}$; (Middle, left) For $a_{1}$ and $a_{4}$; 
(Middle, right) For $a_{2}$ and $a_{3}$; (Bottom, left) For $a_{2}$ and 
$a_{4}$; (Bottom, right) For $a_{3}$ and $a_{4}$.}
\label{corr2} 
\end{figure*}

We also checked the 
correlation of the splitting coefficients derived for $l$=2 .  Figure \ref{corr1} and \ref{corr2} show respectively the variance and 
the covariance of the splitting coefficients as a function of the leakage elements error.
It can be concluded that the splitting coefficients become correlated 
only when a large overestimation of about 50\% is made for the 
off-diagonal leakage elements.  This result is only valid when 
fitting Fourier spectra.  For other methods, such as fitting power 
spectra, possible correlation amongst the splitting coefficients could 
have drastic consequences for the inverted solar rotation profiles.

\section{Conclusion}
We have given a step by step recipe for fitting $(m,\nu)$ Fourier 
spectra.  If one wants to implement similar fitting technique, one 
should compute, first the leakage matrices according to Eqs. 
(\ref{d}) and (\ref{d6}), second the 
mode covariance matrices with Eq. (\ref{g}) (or using Eq. 
(\ref{h})), third the noise covariance matrices with Eq. (\ref{dd}) 
using a model of solar noise, 
fourth compute the likelihood function using Eq. (\ref{f}).  The 
use of Monte-Carlo simulations will ensure the success of the 
implementation.  Routines for fitting p-mode Fourier spectra are 
available as freeware on the VIRGO home page: 
ftp://ftp.estec.esa.nl/pub/loitenerife/html/software.html; they are written in the IDL language.  

Last but not least, Eq. (\ref{j}) showed us the 
equivalence between fitting data for which the leakage 
matrix is not the identity, and fitting data for which it explicitly is.  
This last statement is true provided that we know perfectly 
well the leakage matrix.  In this case the p-mode parameters fitted using MLE 
are not, or very weakly, biased, and have minimum variance.  
We have also studied the effect of an 
imperfect knowledge of the leakage matrix on the fitted parameters, 
in order to derive the effect of systematic errors on the most 
interesting parameters: the splitting and mode frequency.  We found 
that the central frequency is insensitive to systematic errors in the
leakage matrix, while the splitting coefficients ($a_{i}$) have a 
quadratic dependence upon those errors.  These systematic errors will 
have influence on the inverted solar rotation profiles.

Finally, we would like to stress again that the correct statistical 
treatment of the p-mode data is of vital importance for deducing 
unbiased p-mode parameters.

\begin{acknowledgements} 
Many thanks to Takashi Sekii for constructive comments on the 
manuscript, and for extensive cyberspace  chatting :-) .
\end{acknowledgements}

\appendix
\section{Appendix}
The purpose of this appendix is to show that using a proper matrix 
$\tens{C}_{\tens{B}}$, the noise covariance matrix of Eq. 
(\ref{linear}) given by:
\begin{equation}
\tens{B'}^{(l,l')}=\tens{C}_{\tens{B}}^{-1} \tens{B}^{(l,l')} {\tens{C}_{\tens{B}}^{\rm T}}^{-1}
\label{linear1}
\end{equation}
can have a diagonal form.  The matrix $\tens{B}^{(l,l')}$ can be 
diagonalized and we can write.
\begin{equation}
\tens{B}^{(l,l')}=\tens{P}^{-1} \tens{b}^{(l,l')} {\tens{P}^{\rm T}}^{-1}
\label{linear2}
\end{equation}
where $\tens{b}^{(l,l')}$ is diagonal and $\tens{P}$ is an orthogonal matrix ($\tens{P}^{-1}=\tens{P}^{\rm 
T}$).  Replacing Eq. (\ref{linear2}) into Eq. (\ref{linear1}), we have:
\begin{equation}
\tens{B'}^{(l,l')}=\tens{C}_{\tens{B}}^{-1} \tens{P}^{-1} \tens{b}^{(l,l')} {\tens{P}^{\rm T}}^{-1} {\tens{C}_{\tens{B}}^{\rm T}}^{-1}
\label{linear3}
\end{equation}
Since $\tens{B}^{(l,l')}$ is positive definite all its eigenvalues are 
positive, therefore the square root of $\tens{b}^{(l,l')}$ is defined.  
Therefore if we apply the following transformation to the data:
\begin{equation}
\tens{C}_{\tens{B}}=\tens{P}^{-1} \sqrt{\tens{b}^{(l,l')}}
\label{linear4}
\end{equation}
we can rewrite Eq. (\ref{linear3}) as:
\begin{equation}
\tens{B'}^{(l,l')}=\tens{I}
\label{linear5}
\end{equation}
where $\tens{I}$ is the identity matrix.  So replacing $\tens{C}$ in Eq. 
\ref{j2} by $\tens{C}_{\tens{B}}$ will have the effect of removing 
the artificial correlation due to the 
noise, and also of performing a normalization.  We should point out that 
the transformation matrix $\tens{C}_{\tens{B}}$ that can achieve this 
is not unique, and any multiplication by an orthogonal matrix will 
achieve this.  Nevertheless, we give a solution to the problem which 
can be solved as an eigenvalue and eigenvector problem.  The 
transformation given above does not remove the artificial correlation 
due the p modes but more or less preserve it.  This can have some 
useful application when one wants to produce spectra with uncorrelated 
noise but with correlated p-mode signals.

\end{document}